
\overfullrule=0pt
\parindent=24pt
\baselineskip=18pt
\magnification=1200

\def\ni{\noindent}
\def\op#1#2{{\displaystyle#1\above0pt\displaystyle#2}}
\def\rl{\rightline}
\def\cl{\centerline}
\def\via{{\it via}}
\def\gev{{\rm GeV}}
\def\tev{{\rm TeV}}

\def\e6{{\rm E}_6}
\def\su#1{{\rm SU(#1)}}
\def\sua{[{\rm SU(3)}]^3}
\def\sub{{\rm SU(3)}_{\rm C}\times {\rm SU(3)}_{\rm L}\times
            {\rm SU(3)}_{\rm R}}
\def\sm{{\rm SU(3)}_{\rm C}\times {\rm SU(2)}_{\rm L}\times
            {\rm U(1)}_{\rm Y}}
\def\mi{{\rm M}_{\rm I}}
\def\mc{{\rm M}_{\rm C}}

\def\mpl{{\rm M}_{\rm PL}}
\def\mpln{{\rm M}_{\rm PL}=2.4\times10^{18}~ \gev}
\def\lepnonet#1{{\rm L}_#1({\bf1},{\bf3},{\bar{\bf3}})}
\def\mirlepnonet#1{{\bar{\rm L}}_{\rm #1}({\bf1},{\bar{\bf3}},{\bf3})}
\def\quanonet#1{{\rm Q}_{\rm #1}({\bf3},{\bar{\bf3}},{\bf1})}
\def\mirquanonet#1{{\bar{\rm Q}}_{\rm #1}({\bar{\bf3}},{\bf3},{\bf1})}
\def\aquanonet#1{{\rm Q}^{\rm c}_{\rm #1}({\bar{\bf3}},{\bf1},{\bf3})}
\def\miraquanonet#1{{\bar{\rm Q}}^{\rm c}_{\rm #1}({\bf3},{\bf1},{\bar{\bf3}})}
\def\lep#1{{\rm L}_{\rm #1}}

\def\qua#1{{\rm Q}_{\rm #1}}

\def\aqua#1{{\rm Q}^{\rm c}_{\rm #1}}

\def\clep{[l=(\nu,e);~e^c;~H;~H';\nu^c,~N]}
\def\cqua{[q=(u,d);~H_3\equiv D]}
\def\caqua{[q^c=(u^c,d^c); H'_3\equiv D^c]}

\def\NPB#1#2#3{Nucl. Phys. {\bf B#1}, #2 (19#3)}
\def\PLB#1#2#3{Phys. Lett. {\bf#1B}, #2 (19#3)}
\def\PRD#1#2#3{Phys. Rev. D {\bf#1}, #2 (19#3)}
\def\PRL#1#2#3{Phys. Rev. Lett. {\bf#1}, #2 (19#3)}


\def\op#1#2{{\displaystyle#1\above0pt\displaystyle#2}}
\def\via{{\it via\/ }}
\def\gev{{\rm GeV}}
\def\tev{{\rm TeV}}

\def\e#1{{\rm E}_{#1}}
\def\su#1{{\rm SU(#1)}}

\def\sua{[{\rm SU(3)}]^3}
\def\sub{{\rm SU(3)}_{\rm C}\times {\rm SU(3)}_{\rm L}\times
            {\rm SU(3)}_{\rm R}}
\def\sm{{\rm SU(3)}_{\rm C}\times {\rm SU(2)}_{\rm L}\times
            {\rm U(1)}_{\rm Y}}

\def\mi{{\rm M}_{\rm I}}
\def\mc{{\rm M}_{\rm C}}

\def\mpl{{\rm M}_{\rm PL}}
\def\mpln{{\rm M}_{\rm PL}=2.4\times10^{18}~ \gev}
\def\lepnonet#1{{\rm L}_#1({\bf1},{\bf3},{\bar{\bf3}})}
\def\mirlepnonet#1{{\bar{\rm L}}_{\rm #1}({\bf1},{\bar{\bf3}},{\bf3})}
\def\quanonet#1{{\rm Q}_{\rm #1}({\bf3},{\bar{\bf3}},{\bf1})}
\def\mirquanonet#1{{\bar{\rm Q}}_{\rm #1}({\bar{\bf3}},{\bf3},{\bf1})}
\def\aquanonet#1{{\rm Q}^{\rm c}_{\rm #1}({\bar{\bf3}},{\bf1},{\bf3})}
\def\miraquanonet#1{{\bar{\rm Q}}^{\rm c}_{\rm #1}({\bf3},{\bf1},{\bar{\bf3}})}
\def\lep#1{{\rm L}_{\rm #1}}

\def\qua#1{{\rm Q}^{\rm #1}}

\def\aqua#1{{\rm Q}^{\rm c}_{\rm #1}}

\def\clep{[l=(\nu,e);~e^c;~H;~H';\nu^c,~N]}
\def\cqua{[q^a=(u^a,d^a);~H^a_3\equiv D^a]}
\def\caqua{[q_a^c=(u_a^c,d_a^c); H'_{3a}\equiv D^c_a]}

\def\frac#1/#2{\leavevmode\kern.1em
               \raise.5ex\hbox{\the\scriptfont0 #1}\kern-.1em
               /\kern-.15em\lower.25ex\hbox{\the\scriptfont0 #2}}

\def\cent#1{\hbox{\rlap/#1}}
\def\vec#1{{\overrightarrow #1}}

\def\taumu{\tau\rightarrow\mu\gamma}
\def\mue{\mu\rightarrow e\gamma}
\def\taumus{\tau\rightarrow\mu\mu{\bar\mu}}
\def\taumue{\tau\rightarrow\mu e{\bar e}}
{\nopagenumbers
\rl{CTP-TAMU-80/91}
\rl{Nov. 1992}
\vskip .5in
\cl{\bf $\tau\rightarrow\mu\mu{\bar\mu}$ and
        $\tau\rightarrow\mu e{\bar e}$ Decays in
        String Models with $\e6$ Symmetry}
\medskip
\cl{Jizhi Wu, ~Shinichi Urano ~and ~Richard Arnowitt}
\cl{\it Center for Theoretical Physics, Department of Physics}
\cl{\it Texas A\&M University, College Station, TX77843}
\vskip .5truein
\cl{\bf Abstract}
\medskip

A detailed analysis on the rare $\tau$ decay {\it via}
$\taumue$ and $\taumus$
in the string models with $\e6$ symmetry is reported.
It is found that
$\Gamma(\taumue)\sim(6-7)\Gamma(\taumus)$
and these rates are in general about 1000 times less than that of
$\Gamma(\mu\rightarrow e\gamma)$.
It is also found that the out-going muon in $\taumue$ is
almost 100\% right-handed polarized
and the out-going electrons would be predominately parallel to each other.
These decay processes may be accessible at the SSC.
\vfill
\eject
}
\pageno=1
\noindent{\it 1. Introduction\qquad}
The existence of a huge number of the degenerate vacua makes it difficult to
investigate the experimental consequences of the superstring theory.
Nevertheless, progress has already been made in extracting the
phenomenological implications of the heterotic string models [1,2,3],
despite the
lack of a theoretical frame work to determine the true vacuum of the theory.
Here, one takes a less fundamental attitude by imposing the phenomenological
requirements that the theory must reduce to the standard model at low energies
[4].

In this paper, we focus on a class of three-generation heterotic string models
which have a symmetry structure
$$\e6\times{\rm (N=1 ~Supergravity)}\times{\rm (Hidden~ Sector)}\eqno(1.1)$$
at the compactification scale $\mc$ close to the Planck scale $\mpln$.
This class of models arises naturally from the compactifications of the ten
dimensional heterotic string [5] on the Calabi-Yau manifolds [6],
and from the N$=$2 superconformal constructions [7],
which allow for the breaking of the $\e6$ group to $\sua\equiv\sub$
by Wilson loops at $\mc$.
The massless multiplets are then in the {\bf 27}, ${\overline{\bf 27}}$
and singlet
representations of $\e6$. In terms of the $\sua$ quantum numbers, these states
are:
$$\eqalign{{\bf27}_i&=\lepnonet{i}\oplus\quanonet{i}\oplus\aquanonet{i};\cr
{\overline{\bf27}}_j&=\mirlepnonet{j}\oplus\mirquanonet{j}
\oplus\miraquanonet{j},
\cr}\eqno(1.2)$$
where $i$, $j$ are generation indices, i.e., $i=1,2,\cdots,n+3$,
$j=1,2,\cdots,n$ for a three generation model.
The notations for these massless multiplets, in terms of the Standard Model
particle notation, are $\lep{}=\clep$,
$\qua{}=\cqua$ and $\aqua{}=\caqua$, where $l$, $H$, $H'$ and $q^a$ are
${\su2}_{\rm L}$ doublets, $e^c$, $u^c_a$ and $d^c_a$ are
the conjugate singlets,
$D^a$ and $D^c_a$ are the color Higgs
triplets of the {\su5} 5 and $\bar 5$
representations ($a=1,2,3$ is the $\su3_{\rm C}$ index),
$\nu^c$ is a {\su5} singlet and $N$ an $O(10)$ singlet.

The breaking of the $\sua$ into the standard model gauge group can be achieved
by the VEV growth of the standard model singlet fields $\nu^c$ and $N$ at a
lower scale $\mi\op{>}{\sim}O(10^{16})$ \gev.
In order to prevent too rapid decay of the proton,
the structure of the couplings
must preserve matter parity [8] at $\mpl$.
Then, it can be proved [9] that the spontaneous breaking of $\sua$ symmetry to
the standard model $\sm$, triggered by the nonrenormalizable interactions and
the supersymmetry breaking mass of O(1~\tev), occurs at an intermediate scale
$\mi\op{>}{\sim}O(10^{16})$ {\gev} in a way such that the lowest lying extremum
of the
effective potential is the one that preserves {\it simultaneously} both
$\su2\times U(1)$ and matter parity.
This breaking is accomplished in one step in the sense that
$N_i$ and $\nu^c_i$ must grow VEV's at the same time and with the same order
of magnitude O($\mi$).

The following theorem regarding the low energy spectrum below the intermediate
scale was established in Ref. [2]:

\ni{\bf Theorem}\qquad For any string model having the symmetry structure of
Eq.~(1.1) and a matter parity invariance at the Plank scale with
\item{(i)} Breaking of $\e6\rightarrow\sua\equiv\sub$ at a scale
$\mc\sim \mpl$,
\item{(ii)} Intermediate scale breaking of $\sua\rightarrow\sm$
at a scale $\mi$, where $\mpl>\mi>10^{16}$~\gev, triggered by a mass
$m<O(1~{\rm TeV})$, and
\item{(iii)} VEVs of $N_i$, $\nu^c_i$ obeying
$\sum\langle N_i\rangle\langle\nu^c_i\rangle=0$,

\ni then there exist, in addition to the three generations of light states
of the Standard Model, always at least two new non-$E_6$ singlet
light chiral multiplets given by
$$n_1=(N_1+{\bar N}_1)/\sqrt{2}, \qquad
{\hat\nu}^c_2=(\nu^c_2+{\bar\nu}^c_2)/\sqrt{2}, \eqno(1.3)$$
and in some cases four new non-$E_6$ singlet light chiral multiplets, with the
additional possible light state given by
$$n_2=cos\theta N_2+sin\theta {\bar\nu}^c_1, \qquad
 {\bar n}_2=cos\theta {\bar N}_2+sin\theta \nu^c_1.\eqno(1.4)$$
Here $tan\theta=\langle\nu^c_2\rangle/\langle N_1\rangle$,
and a basis has been chosen in generation
space such that $\langle N_i\rangle=\langle N_1\rangle\delta_{i1}$
and $\langle\nu^c_i\rangle=\langle\nu^c_2\rangle\delta_{i2}$.
In addition there may also be a number of light $E_6$ singlet multiplets,
$\phi_a$.

We will investigate the phenomenological consequences of this theorem in this
paper.
In Sec.~{\it 2.}, we discuss the interactions of these new particles with the
standard model particles [10,2] and their implications on rare $\tau$ and $\mu$
decays. Sec.~{\it 3} briefly summarizes the results of Ref. [3]
on the fermion family violating processes $\taumu$ and $\mue$.
Sec.~{\it 4.} represents a detailed analysis on the decays of $\tau$ particle
{\via} $\taumus$ and $\taumue$ where the conclusion is reached that these decay
processes may be accessible at the SSC.
Sec.{\it 5.} is devoted to conclusions.

\noindent{\it 2. Low-Energy Effective Interactions and
$\tau$ and $\mu$ Decays\qquad}
The appearance of the new light particles can be regarded as the low
energy remnants of the Planck scale physics dictated by superstring theory.
Thus their detection, either direct or indirect, could be regarded as the
confirmation of the dynamics induced by superstring theory.
Since these new particles are Standard Model singlets and presumably
have a large mass $\sim{\rm O}(1$~\tev) compared to the Standard Model
particles, they are most likely to be detected indirectly by the new
phenomena not predicted by the Standard Model, but which involve
the Standard Model particles

The interactions of these new particles can be determined by examining the
mass matrix below $\mi$ and the Yukawa and gauge couplings.
One finds then [10,2] from the $({\bf 27})^3$ and $({\bf\overline{27}})^3$
pieces in the superpotential,
$$\eqalignno{
W_{eff}=&(\lambda_{pp'}^{(l)}H'e_p^cl_{p'}
         +\lambda_{pp'}^{(u)}Hq_pu^c_{p'}
         +\lambda_{pp'}^{(d)}H'q_pd^c_{p'})\cr
      &+\{[\lambda_pHl_pn_2+{\bar{\lambda}}_pHl_p{\bar n}_2
           +(\lambda_1n_1+{\hat\lambda}_2{\hat\nu}_2^c)HH']\cr
      &\qquad+(\lambda^{(0)}_{ap}\phi_a^{(0)}Hl_p
           +\lambda^{(e)}_{ap}\phi_a^{(e)}H'H)\cr
      &\qquad+(m_1n_2{\bar n}_2+m_2n_2n_2+m_3{\bar n}_2{\bar n}_2)\cr
      &\qquad+(m_4n_1{\hat\nu}_2^c+m_5n_1n_1+m_6{\hat\nu}_2^c{\hat\nu}_2^c)
             +m_{ab}\phi_a\phi_b\}+W_{seesaw},&(2.1)\cr}$$
and from the gaugino ($\lambda_L^{(-)}$) interaction
$$\eqalignno{{\cal L}_{\rm gaugino}
=&g_LU^\dagger_{p\lambda}l_p\gamma^0[\lambda_{pp'}^{(g)}e_{p'}^cH^\dagger
+{1\over\sqrt2}(sn_1-c{\hat\nu}_2^c){\tilde l}^\dagger_{p=1}\cr
&\quad+(\lambda'n_2+{\bar\lambda}'{\bar n}_2)H^{\prime\dagger}
+H(\lambda n_2^\dagger+{\bar\lambda}{\bar n}_2^\dagger)]+h.c.,
&(2.2)\cr}$$
where $s=\sin\theta$ and $c=\cos\theta$.
In eqs.~(2.1) and~(2.2), the $\lambda$'s are various effective coupling
constants
which are related to the elementary coupling constants by various
unitary transformations onto the light fields.
$\phi_a=(\phi_a^{(0)},\phi^{(e)}_a)$ are the C-odd,
C-even $\e6$-singlet fields that remain light.
The $H^\dagger$, $H^{\prime\dagger}$, ${\tilde l}^\dagger$, etc., are scalar
fields, e.g., ${\tilde l}_{p=1}$ is the slepton partner of $p=1$ lepton,
where $p=1,2,3$ are the three light generations of the Standard Model.

The terms in the parentheses in Eq.~(2.1) are just the
low-energy Standard Model superpotential.
Eq.~(2.2) and the terms in the curly brackets in
Eq.~(2.1) are the interactions involving the new
low-energy particles.
These represent the new physics predicted by the
heterotic string models that obey the Standard Model at low energies.
There are also seesaw masses coming from $\nu_p$-Higgs boson-superheavy
interactions, obtained by integrating out the heavy fields.
The consequences of these terms for neutrino masses and neutrino oscillations
were discussed in Ref.~[1] where it was found that the
experimental smallness of neutrino masses, $m_\nu<{\rm O}(10~{\rm eV})$,
leads to restrictions on Yukawa couplings that naturally explains
the smallness of $m_e/m_\tau={\rm O}(10^{-3})$,
and the $\nu_{e,\mu}\leftrightarrow\nu_\tau$ are the dominant oscillations.
Here we are interested in the implications of these new particles and new
couplings on the lepton family number violating processes such as $\tau$
and $\mu$ lepton decays channels.

The interactions of the new light particles discussed above can
trigger lepton family number violating processes such as
$\tau\rightarrow\mu\gamma$, $\mu\rightarrow e\gamma$,
$\taumus$ and $\taumue$, not predicted
by the Standard Model.
Figure~1 demonstrates how $\tau\rightarrow\mu\gamma$
arises from the $n_1$ and ${\hat\nu}_2^c$ interactions.
There are quite a few sources contributing to these processes:
(i) $\nu^c_2$ and $n_1$ through the gaugino interactions
Eq.~(2.2):
$${\cal L}_{\rm g}
=-c{g_L\over\sqrt2}U^\dagger_{p(-)}l_p\gamma^0
{\hat\nu}_2^c{\tilde l}^\dagger_{p=1}
+s{g_L\over\sqrt2}U^\dagger_{p(-)}l_p\gamma^0
n_1{\tilde l}^\dagger_{p=1},\eqno(2.3)$$
where the unitary matrices $U^\dagger_{p(-)}$ are defined from the projection
of the $\su3_{\rm L}$ gauginos, $\lambda^{(-)}_{\rm L}$, onto the  light lepton
doublets, $\lambda^{(-)}_{\rm L}=l_pU^\dagger_{p(-)}+$heavy fields.
This interaction is always present regardless
what the $\e6$ singlet couplings are.
(ii) $n_2$ and ${\bar{n}}_2$ from $W_{eff}$
$$W_{n_2{\bar n}_2}=\lambda_pHl_pn_2+{\bar\lambda}_pHl_p{\bar n}_2,
\eqno(2.4)$$
which is present if $n_2$ and ${\bar n}_2$ also remain light.
(iii) $\e6$ singlets $\phi_a^{(0)}$ give
$$W_\phi=\lambda^{(0)}_{ap}\phi_a^{(0)}l_pH.\eqno(2.5)$$
These are present provided $\phi_a^{(0)}$ remain light.

An estimate on the relative sizes of these sources is given in Ref.~[3].
Thus for the entire reasonable range of
$\varepsilon\equiv\tan\theta$, i.e.,
$10^{-3}<\varepsilon<1$, the dominant effects come from ${\hat\nu}_2^c$ and
$n_1$ which are always present independent of $\e6$ singlet couplings.
Since the ${\hat\nu}_2^c$ and $n_1$ couplings hold for all models satisfying
the
conditions of the Theorem stated above, these decay processes are essentially a
universal prediction for all phenomenologically acceptable models of this type.
We shall hereafter consider only contributions from ${\hat\nu}_2^c$ and
$n_1$ and for simplicity we shall set their masses to a common value
\def\hm{{\hat m}} $\hm$.

\noindent{\it 3. $\mu\rightarrow e\gamma$ and $\tau\rightarrow\mu\gamma$
Decays\qquad}
These processes were studied in Ref.~[3], we briefly summarize
the results here.
We may parametrize the $\mu\rightarrow e\gamma$ decay by an effective
Lagrangian
$${\cal L}={e\over4m_\mu}F^{\alpha\beta}{\bar\mu}\sigma_{\alpha\beta}
(a_R^{(\mu)}P_R+a_L^{(\mu)}P_L)e+h.c.\eqno(3.1)$$
where $\mu(x)$ and $e(x)$ are the lepton fields.
The coefficients $a_L$ and $a_R$ were determined to be
$$\eqalign{
a_R^{(\mu)}\simeq&{\alpha\over8\pi\sin^2\theta_w}\left[{m_\mu\over\hm}\right]^2
L(x)\left[{m_e\over m_\tau}(1+r^2)^{1/2}\right]\varepsilon^4\cr
a_L^{(\mu)}=&\left[{m_e\over m_\mu}\right]a_R^{(\mu)}\ll a_R^{(\mu)}\cr}
\eqno(3.2)$$
where $x=m_{\tilde e}/\hm^2$ and $L(x)$ is the loop integral
$$L(x)={1\over(1-x)^4}\left[{1\over3}+{1\over2}x-x^2+{1\over6}x^3+x\ln x\right]
\eqno(3.3)$$
and $r\sim3.5$.
The total decay rate is proportional to
$a^2_{(\mu)}\equiv\frac1/2(a^2_L+a^2_R)$.
A similar analysis for the $\tau\rightarrow\mu\gamma$ decay yields
$a_{(\tau)}=(m_\tau^2/m_\mu^2)(\delta^2/\varepsilon)^{-1}a_{(\mu)}$,
where $\delta^2$ is yet another parameter of the $\e6$ model and maybe
determined from the electron and tau lepton mass ratio,
$m_e/m_\tau=(\delta^2/\varepsilon)(1+r^2)^{1/2}$ [1].
The following relation is then obtained
$$B(\tau\rightarrow\mu\gamma)
=\left({m_\tau\over m_\mu}\right)^5\left[\left(m_\tau\over m_\mu\right)^2
(1+r^2)^{-1}\right]\left({\Gamma_\mu\over\Gamma_\tau}\right)
B(\mu\rightarrow e\gamma),\eqno(3.4)$$
which implies
$B(\tau\rightarrow\mu\gamma)\simeq2\times10^{5}B(\mu\rightarrow e\gamma)$.
Thus the theory predicts a definite relation between the two lepton number
violating decays.
As a consequence of the ${\hat\nu}^c_2$ and $n_1$ couplings (which arise from
the gaugino couplings), the out-going leptons would be almost 100\%
right-handed
polarized.
This leads to a characteristic angular distribution of the out-going leptons
relative to the spin of the initial lepton even if the spin of the out-going
lepton is not measured.

\noindent{\it 4. $\taumus$ and $\taumue$ Decays\qquad}
These processes arise when the out-going photon in Figure~1 converts into
an electron-positron pair or a muon-anti-muon pair.
These processes cannot be described by an effective Lagrangian as simple as
Eq.~(3.1) which was obtained utilizing the fact that
all the particles involved, in particular the outgoing photon, are on shell.
Here, the intermediate photon is not on-shell, but we may still use the fact
that the in-coming tau particle and the out-going muon arising from the
gaugino vertex are on shell.
Then the $\tau-\mu-\gamma$ vertex can be written as:
$$J^\alpha(p,q)=-i{eg_2 U_{2(-)}^\dagger U_{3(-)}^\dagger\over64\pi^2\hm^2}
                \left[P_L(\gamma^\alpha F_L+f_L^\alpha)
                 +P_R(\gamma^\alpha F_R+f_R^\alpha)\right],
\eqno(4.1)$$
where $p$ and $q$ are four-momenta for the in-coming $\tau$ and the
out-going $\mu$ respectively, $P_{L,R}=\frac1/2(1\mp\gamma_5)$, and
$$\eqalignno{
F_L=&2m_\mu m_\tau(L_2(x)-L_3(x))
&(4.2a)\cr
f_L^\alpha=&-m_\mu\left[(L_2(x)-{2\over3}L_3(x))p^\alpha
                  +(L_2(x)-{4\over3}L_3(x))q^\alpha\right]
&(4.2b)\cr
F_R=&(L_2(x)-{4\over3}L_3(x))(m_\tau^2+m_\mu^2)+{2\over3}L_3p\cdot q
&(4.2c)\cr
f_R^\alpha=&-m_\tau\left[(L_2(x)-{4\over3}L_3(x))p^\alpha
                  +(L_2(x)-{2\over3}L_3(x))q^\alpha\right]
&(4.2d)\cr}$$
and $L_2$ and $L_3$ are loop integrals defined by
$$L_n(x)=\int_0^1dy{y^n\over1+(x-1)y},\quad n=2,3,\quad
x={m_{\tilde e}^2\over{\hat m}^2}.\eqno(4.3)$$
Notice that $F_R$ and $f_R^\alpha$ are much larger than $F_L$ and $f_L^\alpha$.

The differential decay rate for the process $\tau\rightarrow\mu l{\bar l}$,
where $l$ denotes either electron or muon, is given by
$$d\Gamma(\tau\rightarrow\mu l{\bar l})
={1\over2m_\tau}|{\cal M}|^2{d^3\vec{q}\over2E(2\pi)^3}
  {d^3{\vec{q}}_1\over2E_1(2\pi)^3}{d^3{\vec{q}}_2\over2E_2(2\pi)^3}(2\pi)^4
\delta^4(p-q-q_1-q_2),\eqno(4.4)$$
where ${\cal M}$ is the amplitude.
The total rate can be calculated from integrating over $q$, $q_1$ and $q_2$.
In the rest frame of the in-coming $\tau$ lepton, the total decay rate is
reduced into the following form:
$$\eqalignno{
d\Gamma(\tau\rightarrow\mu l{\bar{l}})=&{m_\tau^5\over(4\pi)^3\hm^4}
\left[{e^2g_L^2U_{2(-)}^\dagger U_{3(-)}^\dagger\over4(4\pi)^2}\right]^2\cr
&|{\cal M}'|^2dE_1dE_2\delta(\cos\beta-f(E_1,E_2))\theta(1-E_1-E_2-m_\mu),
&(4.5)\cr}$$
where all the energy and masses are scaled by $m_\tau$, $\beta$ is the
angle between two out-going $l$ particles, and
$f(E_1,E_2)=(1+2m_l^2-m_\mu^2-2(E_1+E_2)+2E_1E_2)/
(2\sqrt{(E_1^2-m_l^2)(E_2^2-m_l^2)})$.
The prefactor in Eq.~(4.5) can be written as
$$A={m_\tau^5\over(4\pi)^3\hm^4}
\left[{e^2g_L^2U_{2(-)}^\dagger U_{3(-)}^\dagger\over4(4\pi)^2}\right]^2
={1\over16(4\pi)^3}{m_\tau^5\over\hm^4}
{\alpha^2\alpha_2^2(U_{2(-)}^\dagger U_{3(-)}^\dagger)^2},
\eqno(4.6)$$
where $\alpha_2=g_2^2/4\pi$.
The values of quantities in the above expression are
$U_{2(-)}^\dagger\sim U_{3(-)}^\dagger\sim1$,
$\alpha=1/137$ and $\alpha_2=0.03322$. We have
$$d\Gamma(\tau\rightarrow\mu l{\bar{l}})={3.4318\times10^{-11}\over\hm^4}
|{\cal M}'|^2dE_1dE_2,\eqno(4.7)$$
where the three body phase space (the range of $E_1$ and $E_2$) is determined
by the $\delta$- and $\theta$-functions in Eq.~(4.5) to be:
$$\eqalignno{(1+2m_l^2-m_\mu^2-2(E_1+E_2)+2E_1E_2)^2
 \le&4(E_1^2-m_l^2)(E_2^2-m_l^2),&(4.8a)\cr
E_1+E_2+m_\mu\le&1.&(4.8b)\cr}$$

For $\taumue$ decay, we have
$$\eqalignno{|{\cal M}'|^2&={1\over4(p-q)^4}
({\rm tr}{\cent{q}}_1\gamma_\alpha{\cent{q}}_2\gamma_\beta
-m_e^2{\rm tr}\gamma_\alpha\gamma_\beta)\cr
&\quad\left[{\rm tr}(\cent{p}+1)R^\beta\cent{q}R^\alpha
+{\rm tr}(\cent{p}+1)(L^\beta\cent{q}L^\alpha
+m_\mu(R^\beta L^\alpha+L^\beta R^\alpha))\right]&(4.9)\cr}$$
where
$$L^\alpha=\gamma^\alpha F_L+f_L^\alpha,\qquad
  R^\alpha=\gamma^\alpha F_R+f_R^\alpha.\eqno(4.10)$$
After integrating over the phase space, we have
$$\Gamma(\tau\rightarrow\mu e{\bar e})
={1.5501\times10^{-10}\over\hm^4}\left[
(L_3-0.987L_2)^2+1.8237\times10^{-3}L_2L_3\right]\eqno(4.11)$$

Similarly, for $\tau\rightarrow\mu\mu{\bar\mu}$, we have
$$|{\cal M}'|^2={1\over2}\left[{A(q_1,q_2)\over(p-q_2)^4}
  +{A(q_2,q_1)\over(p-q_1)^4}-{B(q_1,q_2)+B(q_2,q_1)\over(p-q_1)^2(p-q_2)^2}
\right],\eqno(4.12)$$
where
$$\eqalignno{
A(q_1,q_2)=&{1\over2}{\rm tr}(\cent{q}-m_\mu)\gamma_\alpha({\cent{q}}_1+m_\mu)
            \gamma_\beta\cr
           &\quad{\rm tr}\left[(\cent{p}+1)(L_2^\alpha P_R+R_2^\alpha P_L)
                  ({\cent{q}}_2+m_\mu)(P_LL_2^\beta+P_RR_2^\beta)\right]
&(4.13a)\cr
B(q_1,q_2)=&{1\over2}{\rm tr}\big[(\cent{p}+1)
            (L_2^\alpha P_R+R_2^\alpha P_L)\cr
           &\quad({\cent{q}}_2+m_\mu)\gamma_\beta(\cent{q}-m_\mu)
            ({\cent{q}}_1+m_\mu)\cr
           &\quad(P_LL_2^\beta+P_RR_2^\beta)\big]
&(4.13b)\cr}$$
where $q_1$ and $q_2$ are the momenta of the out-going muons and $q$ is that
of the out-going anti-muon, and $L_i^\alpha$ and $R_i^\alpha$ are given by
Eq.~(4.10) by replacing $q$ by $q_i$ for $i=1,2$.
After integrating over the phase space, we have
$$\Gamma(\tau\rightarrow\mu\mu{\bar\mu})
={3.8217\times10^{-11}\over\hm^4}\left[
(L_3-0.9235L_2)^2+5.8093\times10^{-3}L_2L_3\right].\eqno(4.14)$$

The decay rates $\Gamma(\tau\rightarrow\mu e{\bar e})$
and $\Gamma(\tau\rightarrow\mu\mu{\bar\mu})$
as functions of $x={m_{\tilde e}}^2/{\hat m}^2$ are listed in
Table~1.
We see that for the physically interesting range of $x$, i.e., $0.01<x<100.0$,
$\Gamma(\tau\rightarrow\mu e{\bar e})$ is about $6\sim7$ times
$\Gamma(\tau\rightarrow\mu\mu{\bar\mu})$.
These decay rates are about 1000 times less than that of
$\Gamma(\mu\rightarrow e\gamma)$,
and therefore have much less chance to be detected.
One nice feature about $\tau\rightarrow\mu e{\bar e}$ is that the out-going
muon
is almost 100\% right-handed polarized which may be used as a distinguishing
feature for its detection.
Figure~2 and Figure~3  are Dalitz plots for the processes $\taumue$ and
$\taumus$ respectively.
Figure~4 draws the differential event rate of $\taumue$ as the function of
the angles between the out-going $e\bar e$ pair (normalized to 1000 events).
One finds that aboout 55~\% of events have the $e\bar e$ pair coming out
with angles less than 5$^\circ$ between them, and about 97~\% are forward
with angles between the $e\bar e$ pair being less than 90$^\circ$.
One concludes than that the out-going $e\bar e$ pair will be predominately
parallel to each other.
Thus, for ${\hm=0.5~\tev}$ and $x=0.04$ (corresponding to a selectron mass
of {100~\gev} which is in accord with the experimental lower bound of
$m_{\tilde e}\op{>}{\sim}65~\gev$ [11]), Eqs.~(4.3), (4.11)~and~(4.14)
imply that ${\rm B}(\taumue)=7.16\times10^{-11}$ and
${\rm B}(\taumus)=9.25\times10^{-12}$.
The SSC with a luminosity of
${\cal L}=10^{33}-10^{34}$~$cm^{-2}s^{-1}$ and a B-production cross section
of $\sigma_{\rm B}\sim0.5~{\rm mb}$ will produce $5\times10^{11}$
to~$5\times10^{12}$ $\tau$ events per year.
Therefore, one would expect to have about 30 to 300 events of $\taumue$
and about 4 to 40 events of $\taumus$ at the SSC per year.
With an event acceptance of 10\% and the cleanness of the three lepton events,
we conclude that the rare $\tau$ decay modes, $\taumue$ and $\taumus$,
may be accessible at the SSC.

\noindent{\it 5. Conclusions\qquad}
In conclusion, string models with $\e6$ symmetry are possible viable models
from
the low-energy phenomenological viewpoints.
The rare $\tau$-decays {\it via} $\tau\rightarrow\mu e{\bar e}$
and $\tau\rightarrow\mu\mu\bar\mu$, which may be accessible at the SSC,
give yet one more channel to confront
these models and hence superstring theory with experiments.
\vskip 10truept
This research is supported in part under National Science Foundation Grant No.
PHY-916593.
We thank Dr.~J.~White and Dr.~T.~Kamon for discussions.


%

\def\NPB#1#2#3{Nucl. Phys. {\bf B#1}, #2 (19#3)}
\def\PLB#1#2#3{Phys. Lett. {\bf#1B}, #2 (19#3)}

\def\PRD#1#2#3{Phys. Rev. D {\bf#1}, #2 (19#3)}
\def\PRL#1#2#3{Phys. Rev. Lett. {\bf#1}, #2 (19#3)}

\vskip 15truept
\noindent{\it References}

\item{[1]} {R. Arnowitt and P. Nath, \PLB{244}{203}{90}.}

\item{[2]} {P. Nath and R. Arnowitt, \PRD{42}{2948}{90}.}

\item{[3]} {R. Arnowitt and P. Nath, \PRL{66}{2708}{91}.}

\item{[4]} {M. Dine, V. Kaplunovsky, M. Mangano, C. Nappi and N. Seiberg,
            \NPB{259}{549}{85}.}

\item{[5]} {D.~J.~Gross, J.~A.~Harvey, E.~Martinec and R.~Rohm,
            \PRL{54}{503}{85}, \NPB{256}{253}{85}, and \NPB{267}{75}{86}.}

\item{[6]} {P.~Candelas, G.~Horowitz, A.~Strominger and E.~Witten,
             \NPB{258}{46}{85}, and in {\sl Symp. on Anomalies, Geometry,
             Topology}, eds. W.~A.~Bardeen and A.~R.~White,
             (World Scientific, Singapore, 1985).}

\item{[7]} {D.~Gepner, \PLB{199}{380}{87}, \NPB{296}{757}{87}.
            For a review, see:
            D. Gepner, Lectures on $N=2$ string theory, {\it in}
            SUPERSTRINGS'89, Proceedings of the Trieste Spring School,
            April, 1989. eds. M. Green, R. Iengo, S. Randjbar-Daemi, E. Sezgin
            and A. Strominger (World Scientific, Singapore, 1990).}

\item{[8]} {M. C. Bento, L. Hall and G. G. Ross, \NPB{292}{400}{87};
            G. Lazarides and Q. Shafi, \NPB{308} {451}{88}.}

\item{[9]} {P. Nath and R. Arnowitt,  \PRD{42}{2006}{89};
            J.~Wu, R.~Arnowitt and P.~Nath, Texas A\&M and Northeastern
            Preprint, CTP-TAMU-75/91 \& NUB-TH.3031-91 (Sept.~1991).}

\item{[10]} {R. Arnowitt and P. Nath, and \PRD{40}{191}{89}.}

\item{[11]} {Particle Data Group: Review of Particle Properties,
            \PRD{45}{11-II, S1-S584}{92}.}

\medskip
\noindent{\it Figure Caption\qquad}
\medskip
\item{Figure 1.}
           Decay $\tau\rightarrow\mu\gamma$ arising from intermediate
           ${\hat\nu}^c_2$ and $n_1$.
           This decay is also induced by $\phi_a$, $n_2$ and
           ${\bar n}_2$ fermions.
           Additional diagrams exist with the photon emitted by the
           initial and final fermions.
           The decays $\tau\rightarrow\mu e{\bar e}$ and
           $\tau\rightarrow\mu\mu\bar\mu$ arise when the out-going
           photon converts into $e$-$\bar e$ pair or $\mu$-$\bar\mu$ pair.
\medskip
\item{Figure 2.} Dalitz plot of $\taumue$ (in the rest frame of $\tau$).
The variables along the two
axises are $(p-q_1)^2$ and $(p-q_2)^2$ where $p$, $q_1$ and $q_2$ are the
momenta of $\tau$ and the out-going $e$ and $\bar e$, respectively.
$x=m_{\tilde e}^2/{\hat m}^2$ is taken to be 1.
\medskip
\item{Figure 3.} Dalitz plot of $\taumus$ (in the rest frame of $\tau$).
The variables along the two
axises are $(p-q_1)^2$ and $(p-q_2)^2$ where $p$, $q_1$ and $q_2$ are the
momenta of $\tau$ and the two out-going $\mu$'s, respectively.
$x=m_{\tilde e}^2/{\hat m}^2$ is taken to be 1.
\medskip
\item{Figure 4.} The differential event rate of $\taumue$ as the function
of the angles between $e\bar e$ pair, normalized to 1000 events, for
$x=m_{\tilde e}^2/{\hat m}^2=1$.
One sees that the $e\bar e$ pair would mostly come out parallel to each other,
e.g., 55~\% of them will have an angle between $e\bar e$ pair less than
5$^\circ$.
\medskip
\medskip
\medskip
\noindent{\it Table 1.\qquad}
{Decay Rates as Functions of $x=m_{\tilde e}^2/{\hat m}^2$,
with ${\hat m}$ in GeV.}
\vskip .2truein
\vbox{\offinterlineskip
      \hrule
      \halign{&\vrule#&
      \strut\hfil#\hfil\hfil\cr
     height2pt&\omit&&\omit&&\omit&&\omit&\cr
     &\qquad$x={m_{\tilde e}^2/{\hat m}^2}$\qquad\hfil&
     &\qquad${\Gamma_{\tau\rightarrow\mu e{\bar e}}\times{\hat m}^4}$\qquad&
     &\qquad${\Gamma_{\tau\rightarrow\mu\mu{\bar\mu}}\times{\hat m}^4}$\qquad&
     &\quad Ratio \quad&\cr
     \noalign{\hrule}
    height2pt&\omit&&\omit&&\omit&&\omit&\cr
      &$0.0001$&&$2.44\times10^{-11}$&
      &$1.51\times10^{-11}$&&$1.61$&\cr
    height2pt&\omit&&\omit&&\omit&&\omit&\cr
      &$0.001$&&$1.81\times10^{-11}$&
      &$6.43\times10^{-12}$&&$2.81$&\cr
    height2pt&\omit&&\omit&&\omit&&\omit&\cr
      &$0.01$&&$1.33\times10^{-11}$&
      &$2.21\times10^{-12}$&&$6.00$&\cr
    height2pt&\omit&&\omit&&\omit&&\omit&\cr
      &$0.1$&&$6.67\times10^{-12}$&
      &$8.41\times10^{-13}$&&$7.93$&\cr
    height2pt&\omit&&\omit&&\omit&&\omit&\cr
      &$1.0$&&$9.91\times10^{-13}$&
      &$1.46\times10^{-13}$&&$6.78$&\cr
    height2pt&\omit&&\omit&&\omit&&\omit&\cr
      &$10.0$&&$3.02\times10^{-14}$&
      &$4.90\times10^{-15}$&&$6.17$&\cr
    height2pt&\omit&&\omit&&\omit&&\omit&\cr
      &$100.0$&&$3.88\times10^{-16}$&
      &$6.40\times10^{-17}$&&$6.05$&\cr
    height2pt&\omit&&\omit&&\omit&&\omit&\cr
      &$1000.0$&&$4.00\times10^{-18}$&
      &$6.65\times10^{-19}$&&$6.03$&\cr
}\hrule}
\vfill
\eject
\end
\bye